**Y-coupled terahertz quantum cascade lasers**

Owen P. Marshall[1], Subhasish Chakraborty[1*], Md Khairuzzaman[1], Harvey E. Beere[2] and David A. Ritchie[2]

[1]School of Electrical and Electronic Engineering, University of Manchester, Manchester, M13 9PL, UK.

[2]Cavendish Laboratory, University of Cambridge, JJ Thomson Avenue, Cambridge, CB3 0HE, UK.

*Email: s.chakraborty@manchester.ac.uk

Here we demonstrate a Y-coupled terahertz (THz) quantum cascade laser (QCL) system. The two THz QCLs working around 2.85 THz are driven by independent electrical pulsers. Total peak THz output power of the Y-system, with both arms being driven synchronously, is found to be more than the linear sum of the peak powers from the individual arms; 10.4 mW compared with 9.6 mW (4.7 mW + 4.9 mW). Furthermore, we demonstrate that the emission spectra of this coupled system are significantly different to that of either arm alone, or to the linear combination of their individual spectra.

In recent years, advancements in terahertz (THz) quantum cascade laser (QCL) technology have greatly improved the viability of these compact, electrically-driven lasers as sources for real-world THz applications. However, one of the remaining barriers to wider adoption of THz QCLs is a lack of electronically controlled single frequency tuning. To date, demonstrations of frequency control in THz QCLs have suffered limitations: active region design (coarse gain tuning) [1-3], waveguide gratings (rigid frequency selection) [4-6], and opto-mechanical control mechanisms (limited switching speeds, complexity of integration) [7, 8]. Many of the photonic engineering strategies used to achieve electronic tuning at shorter wavelengths face practical difficulties when scaled to THz QCLs. For example, the sampled-grating structures described in reference [9] would require a total device length on the order of centimetres if implemented in a 3 THz QCL, a size difficult to fabricate and power. An alternative approach, without the need for very long structures, is to laterally couple closely spaced laser cavities. A variety of lateral coupling schemes have been employed in shorter wavelength solid state lasers (including mid-infrared QCLs) to manipulate output powers, beam shapes and lasing frequencies [10-12], but transference of these ideas to THz QCLs has been very limited. A notable exception was the work of Kao et al, who used metal-metal waveguides with high confinement factors to phase-lock an array of parallel surface-emitting THz QCLs by way of a series of linking phase sectors; coupling waveguide sections with tight 180° bends [13]. However, this structure does not allow for facet emission of individual lasers and the curvatures required are not compatible with the semi-insulting single-plasmon (SI-SP) waveguides often used in THz QCLs. Conversely the SI-SP waveguide permits an alternative coupling mechanism not possible in the metal-metal system. A large fraction of the modal power in a SI-SP QCL lies in the substrate, outside the active region, allowing the substrate to act as a power-coupling channel between adjacent QCL ridges. Total system length may therefore be kept to a minimum by utilizing this lateral coupling mechanism, without the need for sharp waveguide bends.

In this letter we present a laterally-coupled system of two THz QCLs on a single substrate in a Y configuration, differing from earlier mid-infrared Y-junction QCLs in that the two arms were brought into close proximity but not physically merged, allowing each arm to be independently electrically biased. Devices were fabricated from an MBE-grown GaAs/Al$_{0.15}$Ga$_{0.85}$As wafer, V557, with an active region based on reference [14]. Previous V557 QCLs based on standard Fabry-Pérot cavities displayed numerous lasing modes around 2.85 THz, with a trend towards higher frequency modes at higher driving currents/biases due to variation in the 90 active region period lengths; shorter periods possessing higher laser transition energies and alignment biases [15]. The arms of each Y-system were 160 μm-wide, 11.7 μm-tall SI-SP ridge waveguides, defined by wet chemical etching on the same semi-insulating GaAs wafer substrate. Figure 1(a) illustrates the primary dimensions of the ridges: the length, $L_1$ in which the ridge bases were separated by < 10 μm, symmetric arm S-bends with 3 mm radii of curvature, and final separation of 1 mm in length $L_2$. Ohmic contacts were thermally deposited and annealed on the tops of and alongside the ridges (100 nm Pd/Ge and 200 nm AuGeNi respectively), and were subsequently covered with a Ti/Au (20/180 nm) overlayer. The total device length, $L_{tot}$, was measured as the linear distance between (and perpendicular to) the cleaved laser facets. Devices were indium soldered to copper heatsink packages and arms were independently wire-bonded in a two terminal electrical configuration, as shown in Fig. 1(b). Finally, in order to prevent electrical cross-talk between arms a 2 μm-deep by 2 μm-wide trench was focussed ion beam (FIB) milled along the centre line of the devices, breaking the highly n$^+$-doped GaAs layer beneath the QCL active regions. The scanning electron microscope (SEM) image in Fig. 1(c) shows the diverging Y arms, milled trench and ridge bond wires. Lasers were cooled to 4.5 K in a Janis ST-100 continuous-flow helium cryostat and all measurements were performed in pulsed operation (1 % duty cycle, 1 μs pulse length). During Y-system operation the electrical pulsers driving each arm were synchronised. Output power measurements were taken with a large area thermopile detector (calibrated to a Thomas Keating absolute THz power meter), placed inside the cryostat < 1 mm from the widely separated QCL facets, removing the need for collection optics and providing high collection efficiency. High resolution (0.075 cm$^{-1}$) emission spectra were recorded in a nitrogen-purged environment at a fixed alignment using a Bruker Vertex 80 Fourier Transform Infra-red Spectrometer and a QMC helium-cooled bolometric detector. The 10 μm arm separation in $L_1$ was chosen to be significantly sub-wavelength in the material ($\lambda_{mat} \approx 30$ μm) in order to ensure a high degree of coupling in the Y-system. Figure 1(d) shows a simulated cross-sectional mode intensity profile within $L_1$ at a frequency of 2.85 THz, generated with the FIMMWAVE software package. Modal power is present in both ridges along with a shared substrate lobe.

Voltage-current (V-I) and light-current (L-I) data for independently operated arms (*A* and *B*) of a Y-coupled THz QCL (Y1) are presented in Figs. 2(a) and (b). Device Y1 had dimensions of $L_{tot} = 6.23$ mm, $L_1 = 520$ μm and $L_2 = 3.36$ mm. Individual performance characteristics of arms *A* and *B* were similar, having lasing threshold currents $I_{th}$ of 1.12 A and 1.11 A, and peak pulsed output powers $P_{max}$ of 4.7 mW and 4.9 mW respectively. Figure 2(c) contains L-I plots for the Y-system (arms driven synchronously), with arm *B* held at fixed driving currents (given by the correspondingly coloured vertical dashed lines in Fig. 2(b)) while sweeping the driving current of arm *A*. Measured output power values are therefore the combination of the contributions from both arms. As opposed to the linear combination of the individual arm powers expected from completely uncoupled lasers, we observe a marked increase in Y-system emission power due to the higher modal gain possible when both THz QCL active regions are aligned. For example, fixing arm *B* around $P_{max}$ ($I_B = 1.65$ A) and sweeping arm *A* produced a total peak Y-system power of 10.4 mW, 0.8 mW higher than the 9.6 mW sum of the individual arm peak powers. Elevated Y-system powers are more clearly seen in Fig. 2(e), where the initial lasing power of arm *B* has been subtracted. Compared to the output of arm *A* alone (black), additional THz power is observed across almost the entire lasing current range and just below $I_{th}$, the latter due to the presence of sub-threshold gain in *A*, supporting and amplifying the existing modal power of *B*. Equivalent results were observed when the arms were exchanged; *A* held at fixed driving current and *B* swept, as shown in Figs. 2(d) and (f), revealing a power increase of approximately the same magnitude. In each case no significant changes in the V-I characteristics were recorded, confirming good electrical isolation between the arms of Y1. Two further Y-coupled devices were fabricated: Y2 with $L_{tot} = 6.40$ mm and $L_1 = 800$ μm, and Y3 with $L_{tot} = 6.21$ mm and $L_1 = 480$ μm. Elevated Y-system emission powers were observed in both Y2 and Y3, as shown in Figs. 2(g) and (h) respectively. Once again, subtracting the initial lasing power of the fixed arm exposes the increase in system power over the swept arm alone, whilst the V-I of the swept arm remains unchanged.

Evidence of optical coupling in Y1 was also observed in its emission spectra. Figure 3 displays lasing spectra collected from Y1 at a selection of driving currents for arms *A* and *B*, operated alone (upper panels) and synchronously (lower panels). Individually arms *A* and *B* lased on multiple current-dependant modes, with a tendency toward higher frequency modes at higher driving currents. For example, the individual laser spectra recorded just above $I_{th}$ in each arm, shown in Fig. 3(a), display multiple modes. In contrast, the corresponding Y-system spectrum in Fig. 3(d) is predominantly single-moded at 2.87 THz, with a side-mode suppression ratio of 13 dB. Other driving current combinations also gave rise to modified Y-system spectra. When the arms of Y1 were operated around $P_{max}$ both the individual spectra in Fig. 3(b) and the Y-system spectra in Fig. 3(e) exhibited multi-moded behaviour, but the latter did not conform to either arm alone, or to a linear combination of their spectral power distributions. A similar result can be seen in Figs. 3(c) and (f), showing spectra recorded when the arms were driven with high (and differing) pulser currents. In this case, in addition to a modified spectral power distribution we also observe a very small red-shift of the Y-system modes of up to ~1 GHz relative to the closest lasing frequencies in either arm alone.

We conclude that the optical coupling between the two arms provided by the Y-configuration has created a laser system distinct from either arm operated independently. The output powers and emission spectra of this Y-system are not reproducible by a linear summation of the characteristics of its constituent component lasers. This initial demonstration of SI-SP coupling in THz QCLs paves the way for more complex devices in which emission spectra may be varied in a user-defined manner, for instance by combining them with established photonic technologies such as waveguide gratings. Even greater frequency control and flexibility might be achieved by increasing the number of independently driven coupled QCLs.

This work was supported by EPSRC First Grant EP/G064504/1 and partly supported by HMGCC.

**FIG. 1. (a)** Key dimensions of the Y-coupled THz QCL devices. **(b)** Photograph of a fully packaged and wire-bonded device. **(c)** SEM image of a section of the same device, showing the diverging arms, flanking Ohmic contacts, bond wires and FIB-milled trench. **(d)** Simulated cross-sectional mode intensity profile within $L_1$ at a frequency of 2.85 THz. Blue through yellow represents low to high mode intensity. The electric field (vertical component) overlap with the active region in each ridge is 5.8 %.

**FIG. 2. (a)** and **(b)** V-I and L-I characteristics of arms *A* and *B* of device Y1 when operated individually. **(c)** Y-system L-I, sweeping the current in arm *A* while *B* was driven at fixed currents indicated by the dashed lines of corresponding colour in Fig. 2(b). **(d)** Y-system L-I, sweeping the current in arm *B* while *A* was driven at fixed currents indicated by the dashed lines of corresponding colour in Fig. 2(a). **(e)** and **(f)** Data from Figs. 2(c) and (d) after subtracting the initial power of the fixed-current arm. L-I plots of individual arms are included for comparison (black). **(g)** (and **(h)**) V-I and L-I characteristics of one arm of device Y2 (and Y3) operated alone (black), and with the second arm driven at peak output power (red), with an initial power subtraction performed as above.

**FIG. 3. (a)** - **(c)** Pulsed lasing spectra of Y1 recorded from arm *A* (black) and arm *B* (red) alone. **(d)** - **(f)** Y-system (arms *A* and *B* synchronously pulsed) lasing spectra at the same arm driving currents.

[1] G. Scalari, C. Walther, J. Faist, H. Beere and D. Ritchie, Appl. Phys. Lett. **88**, 141102 (2006)

[2] J. R. Freeman, O. P. Marshall, H. E. Beere and D. A. Ritchie, Opt. Express **16**, no. 24, 19830-19835 (2008)

[3] L. Lever, N. M. Hinchcliffe, S. P. Khanna, P. Dean, Z. Ikonic, C. A. Evans, A. G Davies, P. Harrison, E. H. Linfield and R. W. Kelsall Opt. Express **17**, 19926-19932 (2009)

[4] L. Mahler, R. Köhler, A. Tredicucci, F. Beltram, H. E. Beere, E. H. Linfield, D. A. Ritchie and A. G. Davies, Appl. Phys. Lett. **84**, 5446-5448 (2004)

[5] B. S. Williams, S. Kumar, Q. Hu and J. L. Reno, Opt. Lett. **30**, 2909-2911 (2005)

[6] L. Mahler, A. Tredicucci, F. Beltram, C. Walther, J. Faist, H. E. Beere, D. A. Ritchie and D. S. Wiersma, Nat. Photon. **4**, 165-169 (2010)

[7] Q. Qin, B. S. Williams, S. Kumar, J. L. Reno and Q. Hu, Nat. Photon. **3**, 732-737 (2009)

[8] J. Xu, J. M. Hensley, D. B. Fenner, R. P. Green, L. Mahler, A. Tredicucci, M. G. Allen, F. Beltram, H. E. Beere and D. A. Ritchie, Appl. Phys. Lett. **91**, 121104 (2007)

[9] V. Jayaraman, Z. Chuang and L. A. Coldren, IEEE J. Quantum Electron. **29**, 18241834 (1993)

[10] J. E. Ripper and T. L. Paoli, Appl. Phys. Lett. **17**, 371-373 (1970)

[11] E. Kapon, Z. Rav-Noy, L. T. Lu, M. Yi, S. Margalit and A. Yariv, Appl. Phys. Lett. **45**, 1159-1161 (1984)

[12] L. K. Hoffmann, C. A. Hurni, S. Schartner, M. Austerer, E. Mujagić, M. Nobile, A. Benz, W. Schrenk, A. M. Andrews, P. Klang, and G. Strasser, Appl. Phys. Lett. **91**, 161106 (2007)

[13] T. Kao, Q. Hu, and J. L. Reno, Appl. Phys. Lett. **96**, 101106 (2010)

[14] S. Barbieri, J. Alton, H. E. Beere, J. Fowler, E. H. Linfield, and D. A. Ritchie, Appl. Phys. Lett. **85**, 1674 (2004)

[15] J. R. Freeman, C. Worrall, V. Apostolopoulos, J. Alton, H. Beere, and D. A. Ritchie, Photon. Tech. Lett. **20**, 303-305 (2008)

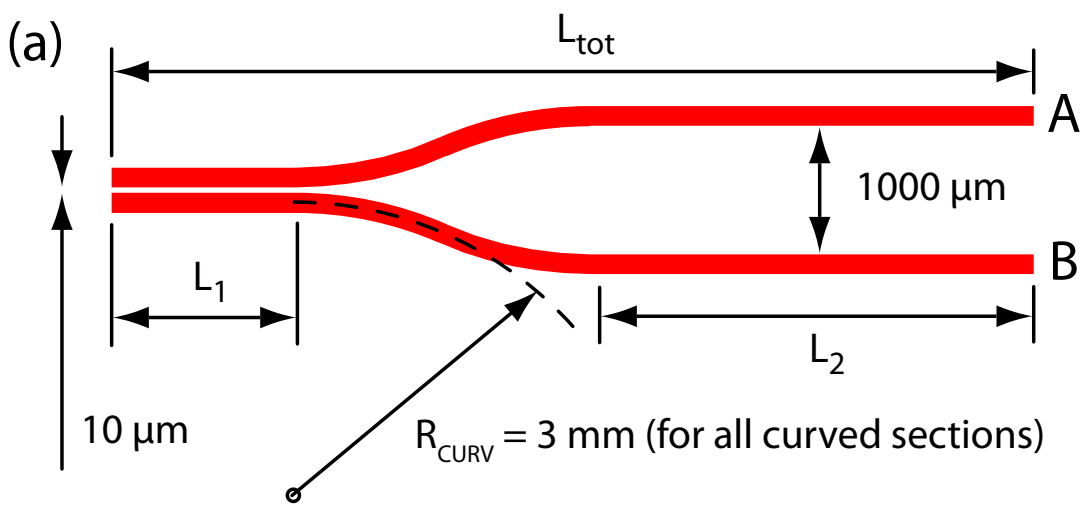
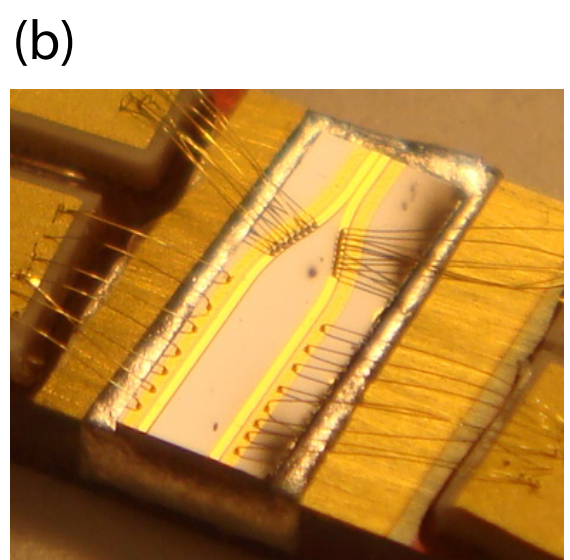
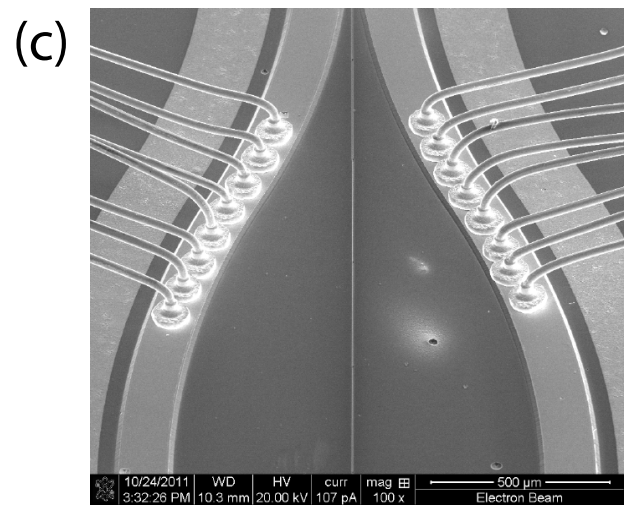
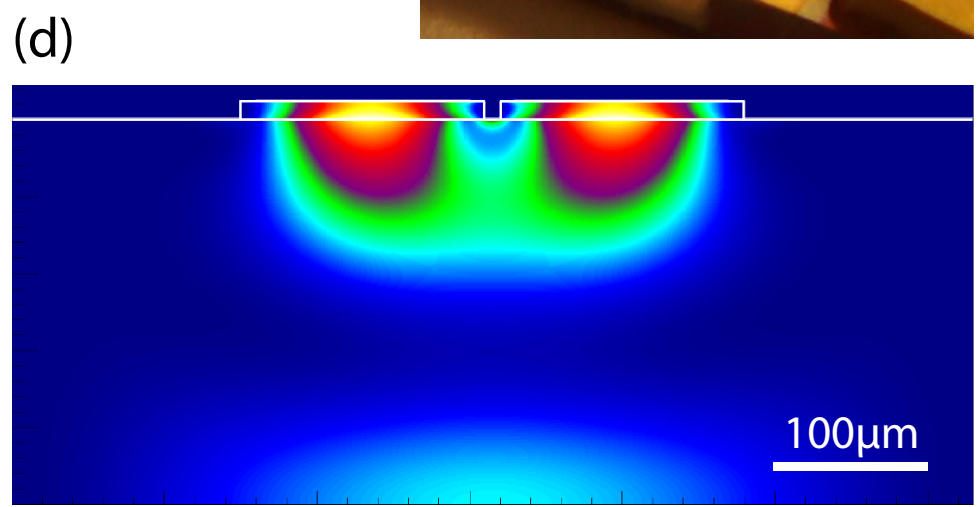

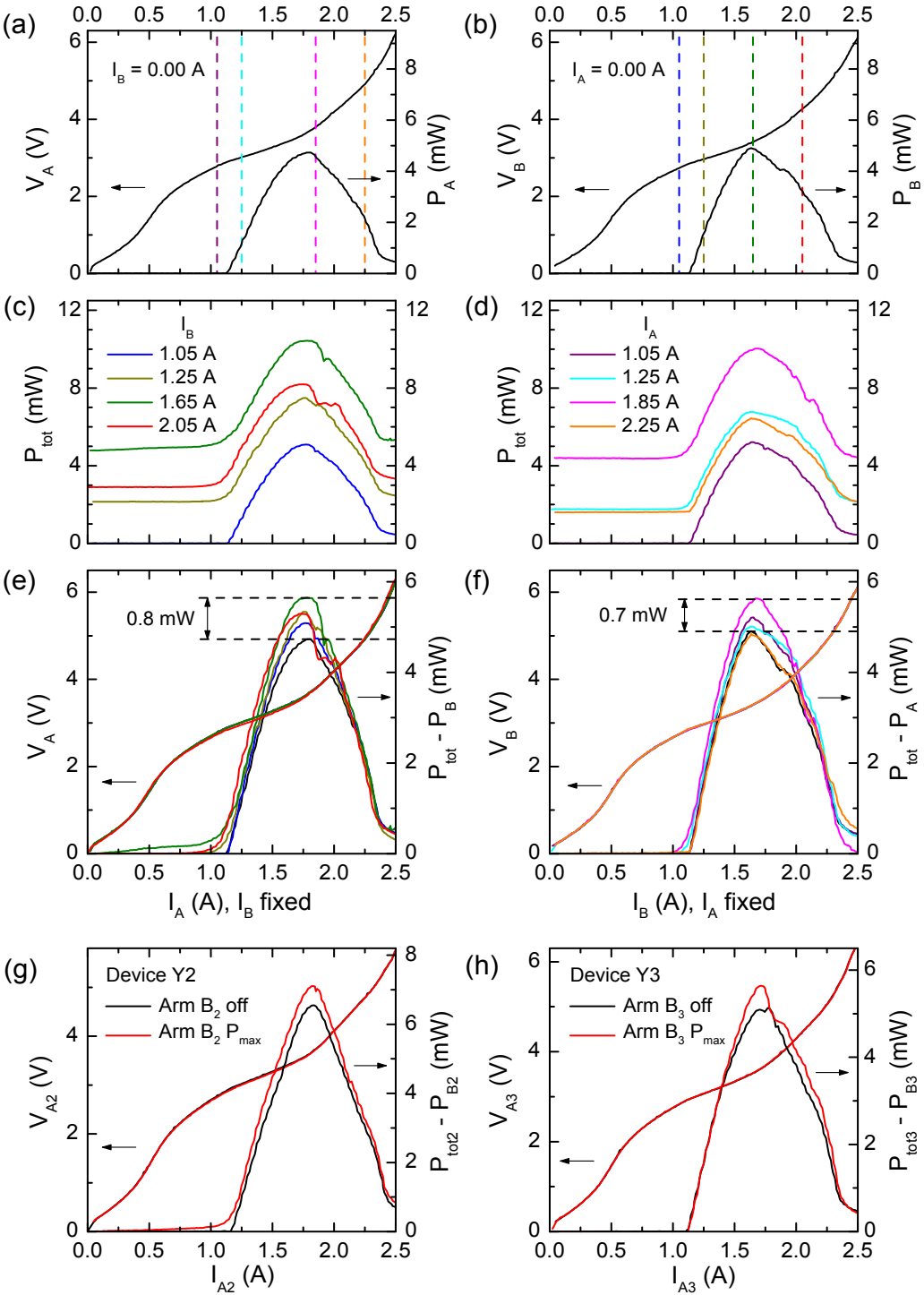

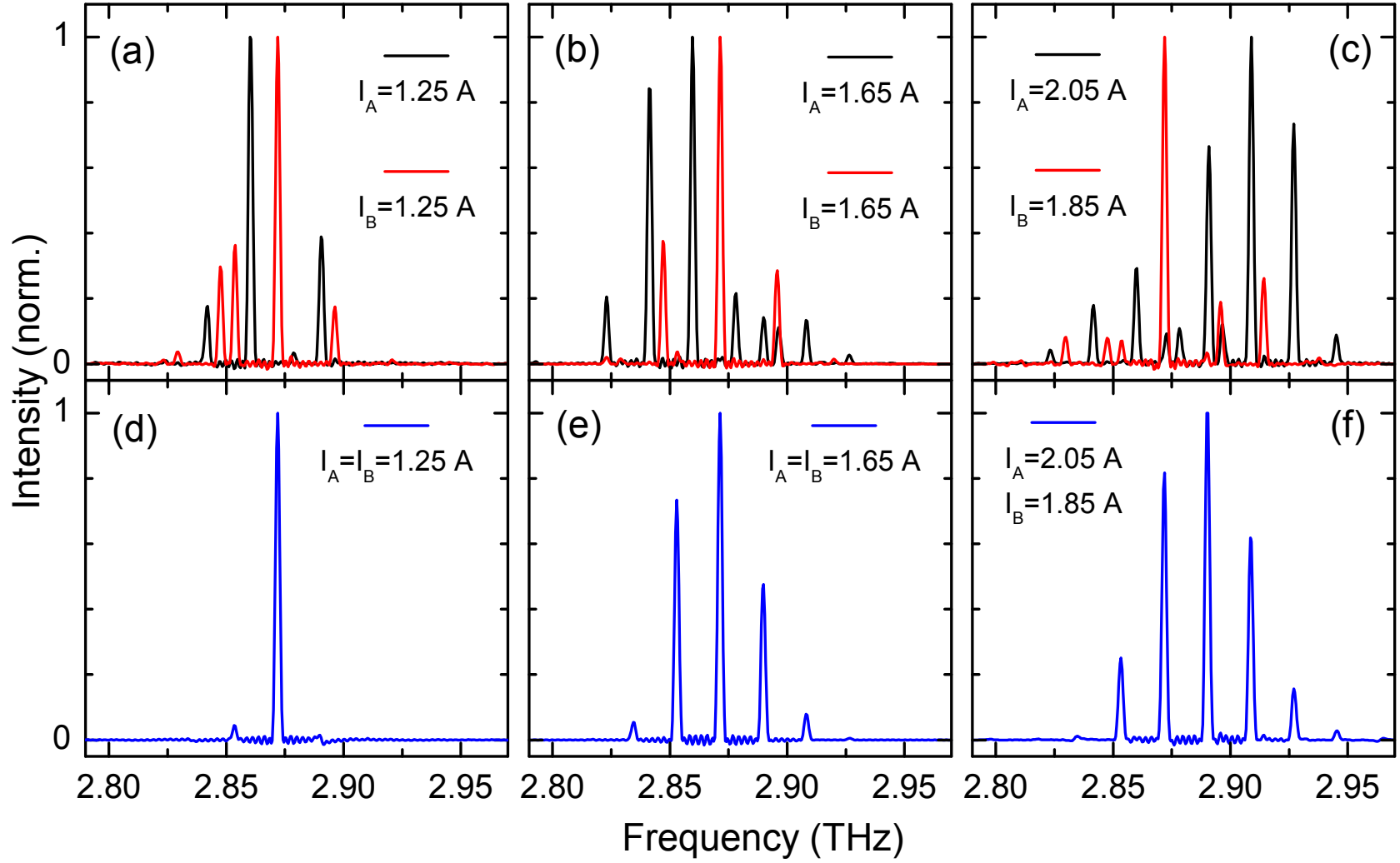